\documentclass[aps,prl,twocolumn,groupedaddress]{revtex4-2}
\usepackage{graphicx,color}
\usepackage{amssymb}   
\usepackage{amsmath}

\usepackage{epstopdf}
\usepackage{natbib}
\usepackage{tabularx,booktabs}
\usepackage[colorlinks=true, letterpaper=true, pdfstartview=FitV, linkcolor=red, citecolor=blue, urlcolor=blue]{hyperref}
\usepackage{bm}
\usepackage{color}
\usepackage{braket}
\usepackage{hyperref}
\usepackage{url}
\usepackage{balance}
\usepackage{xcolor}

\begin{document}
\title{YCl Electride as a Multi-Orbital Correlated Topological Dice Lattice System}
\author{Jianqi Zhong} 
\author{Songyuan Geng} 
\author{Teng-Fei Ying}
\author{Haoxiang Li}
\author{Benjamin T. Zhou} \thanks{Corresponding author: tongz@hkust-gz.edu.cn} 

\affiliation{Thrust of Advanced Materials \& Quantum Science and Technology Center,
The Hong Kong University of Science and Technology (Guangzhou), 1 Duxue Rd., Nansha, Guangzhou, China}

\begin{abstract}
The long-sought dice lattice flat band has recently been discovered for the first time in two-dimensional layered electride yttrium monochloride (YCl) \cite{Geng2026NC}. While essential flat band features of YCl were captured by an idealized simple dice lattice model in Ref.~\cite{Geng2026NC}, we reveal in this Letter that a unique layer-orbital-valley coupling in YCl puts up a fundamental obstruction against a simple three-band dice lattice description of the flat band, and necessitates a multi-orbital description that faithfully represents the symmetry, topology, and correlation physics in the first-ever dice metal. Using an \textit{ab initio}–based multi-orbital Hubbard model with local interactions, we predict that the multi-orbital flat band supports a robust ferromagnetic ground state and electrically tunable correlated quantum anomalous Hall phases that are absent in an interacting single-orbital dice lattice. Our findings open a new avenue for exploring correlation and topology in electride systems.
\end{abstract}

\maketitle
\textbf{}
\emph{Introduction.}--- Interest in searching non-dispersive bands with nontrivial topology has recently been ignited by the discovery of a plethora of correlated topological phases in moir\'{e} flat bands formed by twisted bilayer graphene~\cite{YuanCao2018, Po2018, Liu2019PRB, Song2019PRL, Bultinck2020PRL, Zhang2019PRB, Sharpe2019, Serlin2020, Nuckolls2020}, twisted transition-metal dichalcogenides~\cite{Wu2019PRL, Zhou2022PRR, Zeng2023, Park2023, Goldman2023PRL, Kang2024, Dong2023PRL, Ji2024, Xu2023PRX, Ju2024Review} and their moir\'{e} hetero-bilayers~\cite{Li2021, Zhang2021PNAS, Xie2022PRL, Lu2024, Waters2024, Xie2025NatMater}. However, flat bands realized through moir\'{e} band engineering are generally susceptible to complications due to unavoidable strains and disorders in twist angles~\cite{Lau2022, Kazmierczak2021, Mesple2021, Nakatsuji2022PRB, Tilak2021, Gadelha2021}. Crystalline materials with intrinsic flat bands, while rare in nature, are clearly more desirable in terms of their stability and reproducibility of the flat band physics~\cite{bao2024consecutive, bao2025isolated}.

\begin{figure}[!t]
	\centering
	\includegraphics[width=0.48\textwidth]{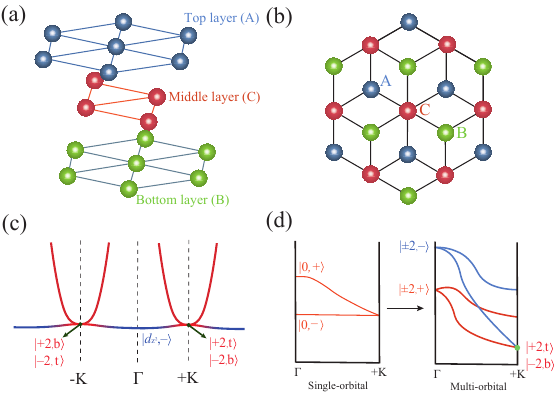}
	\caption{(a) The three-layer effective lattice architecture formed by interstitial anionic electrons(IAEs) within a monolayer YCl. (b) The effective electron lattice has exactly the dice lattice geometry from the top view. (c) Schematic for the band touching with layer-orbital-valley-coupling (LOVC) at $\pm K$ of the flat band, where $\ket{\pm2,l}$ corresponds to $d_{x^2-y^2} \pm i d_{xy}$ orbital in layer $l = t,b$. States near $\Gamma$ are dominated by the anti-bonding state $\ket{d_{z^2}, - }=(\ket{d_{z^2},t}-\ket{d_{z^2},b})/\sqrt{2}$ of $d_{z^2}$-orbital. (d) Band evolutions from $\Gamma$ to $+K$ for single-orbital dice (left panel) and multi-orbital dice model (right panel). $\ket{m_z,\pm}=(\ket{m_z,t}\pm\ket{m_z,b})/\sqrt{2}$ denote the bonding and anti-bonding states with opposite parities for $m_z=0,\pm 2$.}
	\label{fig1}
\end{figure}

In a recent experiment~\cite{Geng2026NC}, the dice lattice and its characteristic non-dispersive band - arguably the first-ever theoretical lattice model proposed to host a non-dispersive band~\cite{Sutherland1986} - has eventually been realized in the two-dimensional electride yttrium monochloride (YCl). Within each quadruple layer unit of YCl - referred to hereafter as a monolayer YCl for simplicity - the Wannier centers of the interstitial anionic electrons (IAEs) are displaced from the ionic cores, forming an effective three-layer architecture (Fig.~\ref{fig1}a). These effective lattice sites constitute the three sublattices A, B, and C of a dice lattice from the top view (Fig.~\ref{fig1}b). Two sets of flat bands with strong band splitting of 0.5 eV were observed in ARPES measurements, and a minimal single-orbital dice lattice model was proposed to explain the essential flat band features~\cite{Geng2026NC}. 

In this Letter, we point out that the idealized three-band dice lattice model in Ref.~\cite{Geng2026NC} is \emph{inadequate} for a faithful representation of the dice lattice flat band in YCl. Due to a unique layer-orbital-valley-coupling (LOVC) in YCl, there exists a fundamental obstruction against a single-orbital dice lattice model that captures all the wave function symmetries in the entire Brillouin zone (BZ), and a multi-orbital description beyond an idealized model is absolutely essential. Importantly, upon introducing interactions, the multi-orbital model predicts electrically tunable correlated spin-polarized topological states, in stark contrast to trivial ferrimagnetic ground states imposed by Lieb's theorem~\cite{liebtheorem1989} on the idealized bipartite dice lattice. These findings establish the YCl electride as a rare natural correlated topological system.

\emph{Symmetry obstruction against single-orbital dice model in the non-interacting limit.}--- The effective lattice architecture of the YCl monolayer, as shown in Fig.\ref{fig1}a,b, respects an overall $D_{3d}$ point group with three generators: (i) three-fold rotation ($\mathcal{C}_{3z}$), (ii) spatial inversion ($\mathcal{P}$), and (iii) mirror reflection ($\mathcal{M}_x: x \mapsto -x$). In the non-relativistic limit without spin-orbit coupling (SOC), the system further respects a spinless time-reversal symmetry $\mathcal{T}$. The IAEs at the A (B) sublattice sites on the top (bottom) layer derive primarily from the 4$d$-orbitals of Y atoms, while the IAEs at the central C site derive from the 5$s$-orbitals (Section I of SM~\cite{SM}). As sublattice A (B) resides in the top (bottom) layer (Fig.~\ref{fig1}a,b), the layer and sublattice degrees of freedom are equivalent; we thus use the terms “layer” and “sublattice” interchangeably in the rest of our discussions.

Although the essential flat band dispersion in YCl is captured by the single-orbital dice lattice model~\cite{Geng2026NC}, the atomic orbital projection at the $+K$ valley reveals significant contributions from $\ket{{+ 2}, t}\equiv d_{x^2-y^2} + i d_{xy}$-orbital in layer $l=t$ (sublattice A) and $\ket{{- 2}, b}\equiv d_{x^2-y^2} - i d_{xy}$-orbital in layer $l=b$ (sublattice B). Following inversion $\mathcal{P}$, the $-K$ valley is dominated by $\ket{{+ 2}, b}\equiv d_{x^2-y^2} + i d_{xy}$-orbital in layer $l=b$ (sublattice B) and $\ket{{- 2}, t}\equiv d_{x^2-y^2} - i d_{xy}$-orbital in layer $l=t$ (sublattice A) (Fig.~\ref{fig1}c). We refer to this unique property of YCl as layer–orbital–valley-coupling (LOVC).


We prove by counter-argument how the LOVC leads to obstruction against a $D_{3d}\otimes \mathcal{T}$-symmetric representation of the flat band with single Wannier functions on A, B, and C sites on a dice lattice (Fig.~\ref{fig1}a). Suppose that one is able to construct single Wannier functions $\ket{\mathcal{W}_{l}}$ on $l = A, B, C$ sites, since both A and B sites serve as rotation centers of $\mathcal{C}_{3z}$, this forces the Wannier functions localized at A and B to be eigenstates of $\mathcal{C}_{3z}$ with eigenvalues $\lambda_m = e^{-i\frac{2\pi  m_{l}}{3}}$, $m_{l=A,B}$ is the magnetic quantum number of $\ket{\mathcal{W}_{l}}$ linked to A and B sites. Since $\mathcal{P}$ swaps the sublattice sites: A $\leftrightarrow$ B, and $\mathcal{T}$ is a \emph{local} symmetry that flips the sign of $m_{l}$, the overall effect of $\mathcal{PT}$ enforces $m_{A} = -m_{B}$. 

The idealized dice lattice model proposed by Sutherland assumes featureless orbitals with $s$-symmetry on sublattice sites with $m_A = -m_B = 0$, thus automatically fulfilling the condition imposed by $\mathcal{PT}$. Due to the LOVC, however, we necessarily have $m_A - m_B = +4$ at $+K$, whereas $m_A - m_B = -4$ at $-K$. As the difference between magnetic quantum numbers cannot be gauged away, it is impossible to construct single Wannier functions on A and B with globally defined $m_A = -m_B$ that correctly reproduce the LOVC effect. This explicitly reveals the obstruction against a single-orbital dice lattice model description of the flat band. 

The obstruction becomes more serious if we compare the symmetry-enforced band degeneracies between a single-orbital dice model and a real band with LOVC. As shown in Fig.~\ref{fig1}d, the band degeneracy at $\pm K$ in the single-orbital dice lattice stems from merging two \emph{non-degenerate} bands with opposite spatial parities at the $\mathcal{P}$-invariant $\Gamma$-point  (left panel of Fig.3d). In contrast, due to LOVC, each component of the doublet $\ket{{+2,t}}, \ket{{-2,b}}$ at $+K$ are formed by \emph{equal superpositions} of bonding and anti-bonding states with opposite parities and $m_z$: $\ket{{+2,t}} = (\ket{{+2,+}} + \ket{{+2,-}})/\sqrt{2}$, $\ket{{-2,b}} = (\ket{{-2,+}} - \ket{{-2,-}})/\sqrt{2}$. Note that $\mathcal{PT}$ requires that $\ket{{\pm2,+}}$ and $\ket{{\pm 2,-}}$ constitute two-fold degenerate time-reversal partners at $\Gamma$, while states with opposite parities can differ in energy at $\Gamma$ (Fig.~\ref{fig1}d). Therefore, the doubly degenerate band at $+K$ will have to evolve and merge into two pairs of doubly degenerate bands at $\Gamma$, in contrast to the two non-degenerate bands in the single-orbital description. This discrepancy reaffirms the necessity of a multi-orbital description.

\emph{Non-trivial band topology in the non-interacting limit.}---The requirement of a multi-orbital description typically enriches the topological and correlated physics. A celebrated example of this is twisted bilayer graphene, where the obstruction against a simple two-band description turns out to render non-trivial flat band topology~\cite{Ahn2019PRX,Po2018,Serlin2020}. As we demonstrate below, the LOVC at $\pm K$ in YCl, which causes the obstruction against a simple three-band description, also endows the multi-orbital flat band with intrinsic nontrivial topology in the non-interacting limit. 

To capture the degeneracies at the Dirac points $\pm K$, we first apply symmetry principles to derive an effective low-energy model in the vicinity of $\pm K$. Under the $D_{3d}\otimes \mathcal{T}$ symmetries, the system respects a particular anti-unitary magnetic point group symmetry $\mathcal{C}_{3z}\mathcal{P} \mathcal{T} \equiv \mathcal{C}_{6z} \mathcal{T}$, which leaves the $\pm K$ points invariant and satisfies $(\mathcal{C}_{6z} \mathcal{T})^2 = \mathcal{C}_{3z} \neq 1$. With $\mathcal{C}_{3z} \equiv e^{i \frac{2\pi}{3} \tau_z}$ and $\mathcal{PT} \equiv \tau_x \mathcal{K}$ in the basis of $\{\ket{{+2}, t}, \ket{{-2}, b} \}$ at $+K$ ($\{\ket{{+2}, b}, \ket{{-2}, t} \}$ at $-K$), the symmetry constraints from $\mathcal{C}_{3z}$, $\mathcal{PT} $ read: $e^{i \frac{2\pi}{3} \tau_z} H_{\pm K}(\bm{p}) e^{-i \frac{2\pi}{3} \tau_z} = H_{\pm K}(\mathcal{C}_{3z}\bm{p})$ and $\tau_x  H^{\ast}_{\pm K}(\bm{p}) \tau_x = H_{\pm K}(\bm{p})$ with momentum $\bm{p} \equiv \bm{k} - (\pm\bm{K})$, leading to a pair of massless Dirac fermions at the $\pm K$ valleys:
\begin{equation}\label{eq:masslessDiracK}
H_{\pm K}(\bm{p}) = v (p_x \tau_x + p_y \tau_y),
\end{equation}

It is worth noting that $\tau_x, \tau_y$ in Eq.~\ref{eq:masslessDiracK} act on \emph{different} layer-orbital subspace at the two different valleys due to LOVC. Thus, the relative chirality between Dirac points at $+K$ and $-K$ is ill-defined, as a local change of basis within one valley does not affect the chirality of the Dirac fermion at the other. This reveals the obstruction due to LOVC in YCl to be \emph{fundamentally} different from that in twisted bilayer graphene, where the chirality of Dirac points at $K$ corners of the moir\'{e} BZ is well-defined~\cite{Po2018,Ahn2019PRX}. 

\begin{figure}[!t]
    \centering
    \includegraphics[width=0.48\textwidth]{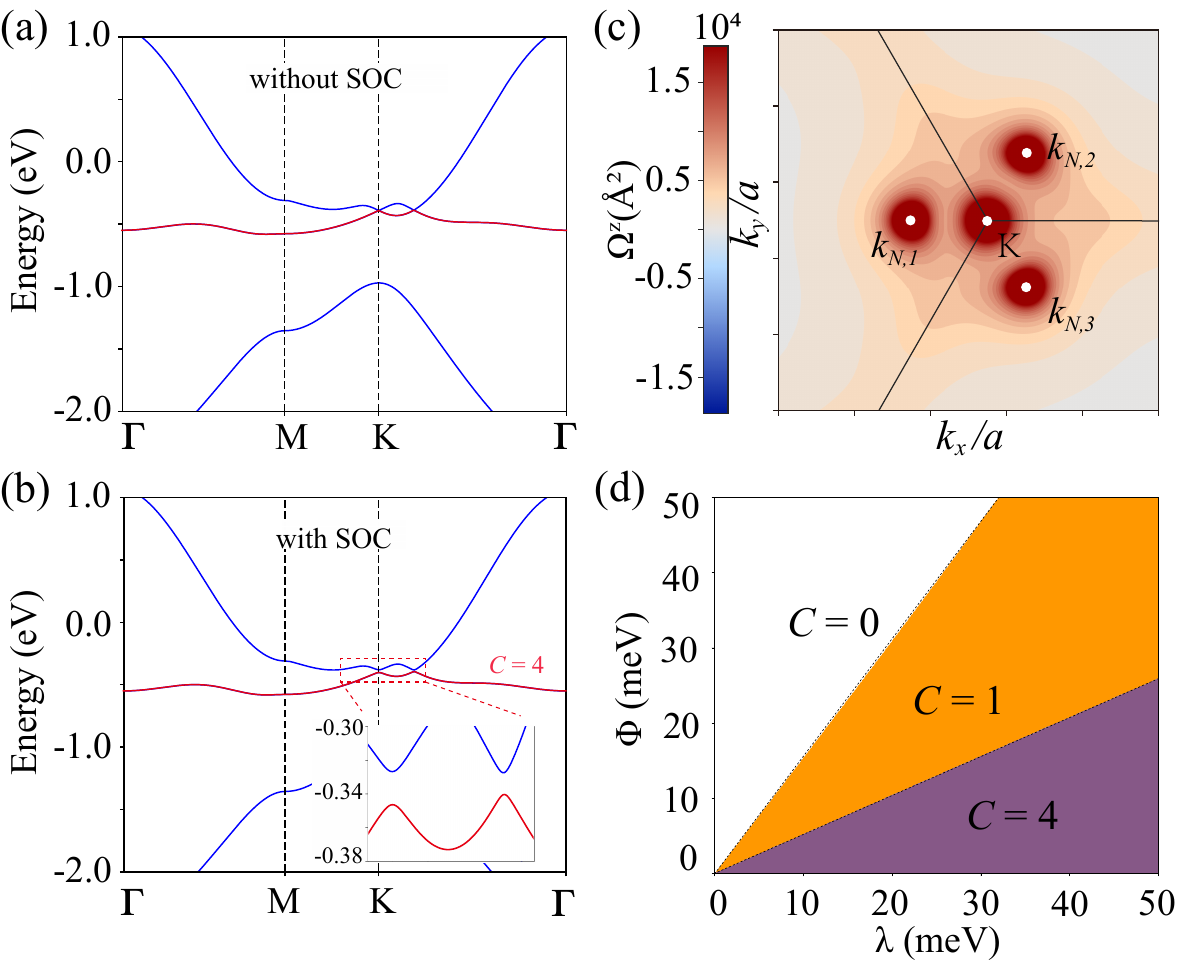}
    \caption{Flat band topology in monolayer YCl. The flat bands in (a)-(b) are highlighted in red. (a)-(b) Band structures obtained using an \emph{ab initio}-based seven-band tight-binding model with atomic SOC strength (a) $\lambda = 0$ in non-relativistic limit, and (b) $\lambda = 15$ meV obtained with relativistic DFT calculations~\cite{SM}. In-set of (b): zoom-in energy bands near Dirac crossing points at $+K$. The topological gap generated by atomic SOC is found to be 20 meV in agreement with DFT calculations (Section I of SM~\cite{SM}). (c) Berry curvature profile of the flat band under $\lambda = 15$ meV in the neighborhood of $+\bm{K}$. Near $+K$ point, 4 Berry curvature hot spots are found at $+\bm{K}, \bm{k}_{\rm N,1}, \bm{k}_{\rm N,2}, \bm{k}_{\rm N,3}$, indicating the presence of 4 massive Dirac fermions. (d) Entire non-interacting topological phase diagram under realistic ranges of $\lambda$ and $\Phi$. } 
    \label{fig2}
\end{figure}

In addition to $\pm K$ points, there also exist other band crossings along the $K' - \Gamma-K$ line (Fig.~\ref{fig2}a-c). As we explain in the SM, it is possible to construct suitable basis states such that the low-energy Hamiltonian near the nodes $\bm{k}_{\rm N}$ is described by massless Dirac fermions:$
H_{N}(\bm{p}) = \tilde{v}_x p_x \tilde{\tau}_z + \tilde{v}_y p_y \tilde{\tau}_y$,    
where $\bm{p} = (p_x, p_y)$ is the momentum measured from $\bm{k}_{\rm N}$, and $\tilde{\tau}_{\alpha=x,y,z}$ are the Pauli matrices under properly chosen bases. The $\mathcal{C}_{3z}$ and $\mathcal{P}$ symmetries guarantee the existence of six Dirac crossing points, hence there are $8 = 2 \times 3 + 2$ massless Dirac fermions in total in the entire BZ. 


The flat band can acquire a global nontrivial topology if proper mass terms are introduced to gap out all the massless Dirac fermions. A prototypical Dirac mass term is provided by atomic spin-orbit coupling(SOC) in the relativistic limit, which is non-negligible for 4$d$ electrons of rare-earth Y elements. In particular, the atomic SOC within the subspace spanned by the basis states $\{\ket{{+2}, \uparrow}, \ket{{-2}, \uparrow}, \ket{{+2}, \downarrow}, \ket{{-2}, \downarrow} \}$ takes the form of:
\begin{equation}\label{eq:SOC}
H_{\rm SOC} = \lambda s_z \tau_z,    
\end{equation}
where $\lambda \approx 15$ meV is the atomic SOC strength obtained with relativistic DFT calculations, $s_z$ denotes the Pauli matrix for spin and $\tau_{z}$ the Pauli matrix in the basis $\{\ket{{+2}}, \ket{{-2}}\}$. Note that only the $s_z$-component is non-vanishing since $\braket{{\pm2}|L_{x,y}|{\pm2}}=0$. The atomic SOC thus reduces to an effective orbital Zeeman field that breaks spinless $\mathcal{T}$ symmetry and splits the doublet $\{\ket{{+2}}, \ket{{-2}}\}$. Therefore, the massless Dirac fermions at $\pm K$ and the six inequivalent $\bm{k}_{N}$ acquire Dirac masses, each contributing to a Berry phase of $\pi$ and summing up to a Chern number of $\mathcal{C} = \pm 4$. We confirm this numerically through an \emph{ab initio}–based seven-band model with atomic SOC in Eq.~\ref{eq:SOC} (Section II of SM~\cite{SM}). As a direct verification of the nontrivial topology, we explicitly calculate the edge spectral function for a YCl monolayer in an infinite-strip geometry (Fig.S2a in SM~\cite{SM}), which explicitly shows 4 branches of propagating chiral edge states.


Thanks to the LOVC, the non-interacting band topology can be effectively tuned by an out-of-plane displacement field $D_z$: as the doublets at $\pm K$ are formed by states from different layers (Fig.~\ref{fig1}), a finite $D_z$ introduces inter-layer potential difference $\Phi$ which provides extra Dirac masses at $\pm K$ with a valley-contrasting sign: $\Delta_{\pm K} = \pm \Phi/2$. As $\Phi \rightarrow \infty$, $\Delta_{\pm K}$ dominates the Dirac mass, and the valley-contrasting Berry curvatures cancel out, making the system topologically trivial. The $\lambda$–$\Phi$ phase diagram (Fig.~\ref{fig2}d) highlights the competition between SOC and displacement field - for larger $\lambda$, a larger $D_z$ is required to drive the system into the trivial phase. 

\emph{Interacting Hamiltonian and Magnetic Phase Diagram.}--- Having established the symmetry and topological properties of the non-interacting flat band, we are introduce interactions. For concreteness, we consider a multi-orbital Hubbard model with on-site interactions:
\begin{equation}
\begin{aligned}
    \hat{H} = & \hat{H}_{0}+ \sum_{i,m}  U_m \hat{n}_{i,m \uparrow} \hat{n}_{i,m \downarrow}+ \sum_{i,m \neq m^{\prime}}  U^{\prime}\hat{n}_{i,m \uparrow} \hat{n}_{i,m^{\prime} \downarrow} \\
    +& \sum_{i,m\neq m^{\prime}}\left(U^{\prime}-J\right) \hat{n}_{i,m \uparrow} \hat{n}_{i,m^{\prime} \uparrow}.
\end{aligned}
\label{Eq4}
\end{equation}
Here, $\hat{H}_{0}$ is the non-interacting seven-band lattice Hamiltonian used to study the nontrivial topology in the previous section. The number operator $\hat{n}_{i,m\sigma}= c^{\dagger}_{i,m\sigma}c_{i,m\sigma}$ counts the density of electrons in orbital $m$ and spin $\sigma = \uparrow, \downarrow$ at $i = A,B,C$ site. $U$ denotes the intra-orbital Coulomb repulsion, $U^\prime$ is the inter-orbital interaction and $J$ is the Hund's coupling among different $d$-orbitals.

\begin{figure}[!t]
	\centering
	\includegraphics[width=0.48\textwidth]{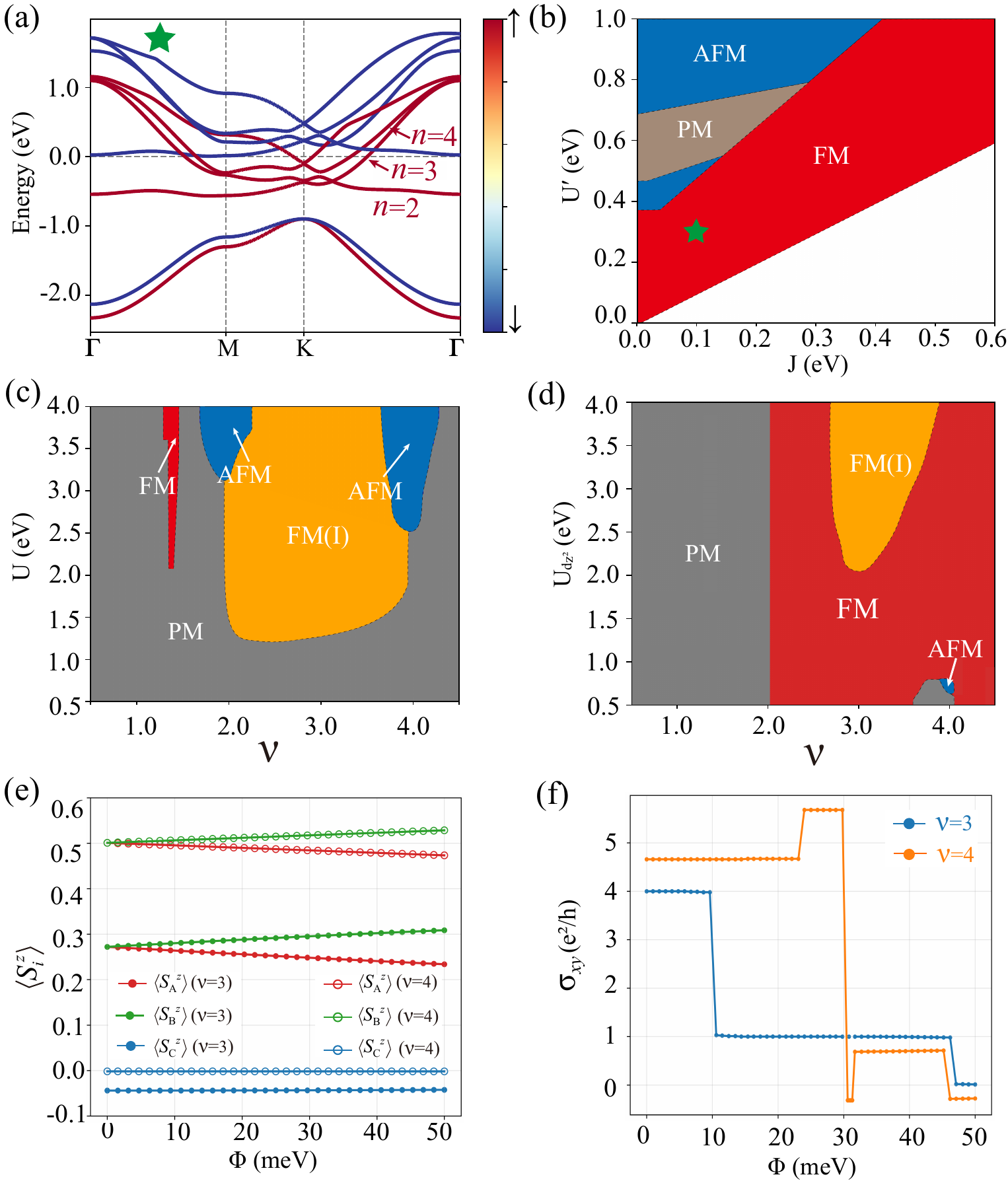}
	\caption{(a) Spin-resolved HF bands under the FM ground state for natural band filling $\nu\simeq4$ with $U^\prime=0.3$ and $J=0.1$ (marked by the green star in panel (b)). (b) Magnetic phase diagram as a function of $U^\prime$ and $J$ at $\nu=4$. Parameters used: $U=2.5$ eV for $d_{x^2-y^2}\pm i d_{xy}$ orbitals, ${U}_{d_{z^2}}=0.9$ eV for $d_{z^2}$ orbital and $U_s=0.3$ eV for $s$-orbital. Parameter choices in (a)(b) are bench-marked from the real band structures observed in the ARPES experiment~\cite{Geng2026NC}. (c)-(d) The magnetic phase diagrams of single-orbital dice lattice (c) and multi-orbital dice lattice (d) as a function of on-site Hubbard $U$ (${U_{d_{z^2}}}$ in multi-orbital model) and doping $\nu$. AFM, FM(I), and PM stand for anti-ferromagnetism, ferrimagnetism, and paramagnetism, respectively. FM(I) phase is identified numerically when the anti-parallel spin moment on C-sites becomes comparable to that on the A/B-sites, with $|\langle S_C^z \rangle| > 0.08$. (e) Spin-magnetic order parameters $\langle S^z_i\rangle$ on sublattice sites $i = A,B,C$ as a function of $\Phi$ at filling factors $\nu=3$ and $\nu=4$. The magnetic orders are largely insensitive to $\Phi$. (f) Self-consistent Hall conductance $\sigma_{xy}$ under HF approximation as a function of $\Phi$ at $\nu=3$ (blue) and $\nu=4$ (orange). 
	}  
	\label{fig3}
\end{figure}

The interacting Hamiltonian Eq.~\ref{Eq4} is solved using the self-consistent Hartree-Fock (HF) method at integer filling factor $\nu=4$~\cite{NoteBandFilling} at $T\approx10$K (Section IV of SM~\cite{SM}). The correlated band structure is shown in Fig.~\ref{fig3}a, which finds excellent agreement with the energy bands observed in ARPES measurements~\cite{Geng2026NC}. The phase diagram in Fig~\ref{fig3}b exhibits three competing magnetic phases, among which the spin-polarized ferromagnetic (FM) state occupies the vast majority of the phase diagram (Fig.~\ref{fig3}b).

The emergence of a dominant FM phase can be understood as a combined effect of two different mechanisms. First, the flat band gives rise to a large density of states at the Fermi level, which fulfills the Stoner criterion ($U_{\text{eff}} \cdot D(E_F) \gtrsim 1$) and drives the instability toward flat-band ferromagnetism. Second, the multi-orbital nature of the flat band imposed by LOVC enables effective direct exchange via on-site inter-orbital Hund's coupling $J$. While the $d_{z^2}$-orbital dominates near the $\Gamma$ point of the flat band, the presence of nearly degenerate $d_{xy}, d_{x^2-y^2}$ states near $\pm K$ allows electrons to occupy these orbitals while retaining parallel spin alignments to save direct exchange energy. 

It is important to note that the correlated physics from a multi-orbital dice model also stands in stark contrast to an idealized single-orbital model, as we demonstrate explicitly in the two contrastingly different $U-\nu$ magnetic phase diagrams in Fig.~\ref{fig3}c,d ($\nu$: band filling factor). In the single-orbital bipartite dice model(Fig.~\ref{fig3}c), where only a local intra-orbital Hubbard $U$ is active on the Y site, the flat-band physics is governed exclusively by lattice connectivity and sublattice imbalance dictated by Lieb's theorem. Consequently, when the flat band is partially filled, the system favors a ferrimagnetic phase (FM(I)) with local spin moments on C sites being anti-parallel to spins on A,B sites. While the quenched kinetic energy enhances interaction effects, the spin configuration follows the geometric constraint rather than spontaneous symmetry breaking. In the multi-orbital description, however, the on-site Hund's coupling and the multi-orbital direct exchange drastically alter the energetics. The fully spin-polarized FM ground state is generally favored (red area in Fig.~\ref{fig3}b,d), and the ferrimagnetic phase with small anti-parallel spin moment on the central C site emerges only under strong $U_{d_{z^2}} > 2$ eV (orange area in Fig.~\ref{fig3}d).

\emph{Electrically Tunable Correlated Topological Phases.}---The dominant FM phase found in our HF calculations (Fig.~\ref{fig3}) suggests that the electron interactions generally lead to strong spin splitting between flat bands with opposite spin polarizations and opposite Chern numbers. At band filling $\nu=3$ where only one of the flat Chern band is fully filled, the system is expected to enter a correlated qauntum anomalous Hall (CQAH) state. As a displacement field effectively tunes the non-interacting band topology (Fig.~\ref{fig2}d), the CQAH phase is also expected to be tunable by gating.

We first demonstrate that the FM order is stable under the displacement field $\Phi$. We define the local spin magnetic order as $\langle S^z_i\rangle = (\langle n_{i,\uparrow} \rangle - \langle n_{i,\downarrow} \rangle)/2$, where $i$ labels the sublattice index. As shown in Fig.~\ref{fig3}e, the FM state remains robust as $\Phi$ increases. With the self-consistent mean-field values of $\langle S^z_i\rangle$ obtained through the HF method, the correlated Hall conductance $\sigma_{xy}$ as a function of $\Phi$ at $\nu =3$ and $\nu =4$ are calculated and shown in Fig.~\ref{fig3}f. It is evident that the correlated $\sigma_{xy}$ at $\nu=3$ closely follows the non-interacting Chern number for the spinless flat band (Fig.~\ref{fig2}d), which confirms that the CQAH phase is electrically tunable. At $\nu=4$, due to the complicated anti-crossings between bands $n=3$ and $n=4$ in Fig.~\ref{fig3}a, $\sigma_{xy}$ is not exactly quantized (yellow curve in Fig~\ref{fig3}f), while giant correlated anomalous Hall effect still remains and exhibits strong dependence on $\Phi$. Details on the dramatic behavior in $\sigma_{xy}$ for $\Phi \approx 20$–$35$ meV at $\nu=4$ are discussed in Section V of SM~\cite{SM}. 

As the range of $D_z$ needed to drive the system across different topological regimes is well accessible in state-of-the-art dual-gate experiments, the correlated QAH phase and anomalous Hall signals can be easily tuned by electrostatic gating under realistic settings, which implies potential device applications in topological electronics~\cite{Gilbert2021}.

\emph{Conclusion and Outlooks.}--- Our findings establish YCl as the first electride that hosts intrinsic correlated topological phases, opening a new avenue for exploring exotic correlated electrons in electride systems-a realm previously studied only within quantum chemistry. Different from early theoretical proposals of topological bands in single-orbital toy dice lattice models where nontrivial topology requires extrinsic symmetry breaking~\cite{Wang2011PRB, Shen2025arXiv}, the correlated topological phase in YCl emerges from its unique intrinsic LOVC effect, which plays the central role behind all interesting physics such as symmetry obstruction, the atomic SOC, the nontrivial band topology and the correlation physics.

The highly non-uniform Berry curvature within the flat band (Fig.~\ref{fig2}) makes it far from resembling the lowest-Landau level - a condition thought to be undesirable for fractional Chern insulator (FCI) ground states at fractional band filling~\cite{Jackson2015, Claassen2015PRL, Mera2021PRB, Varjas2022, Ledwith2020PRR}. However, recent theoretical advances suggest generalized trace conditions of quantum geometric tensors ~\cite{Ledwith2023PRB, Liu2025PRX}, which hints at possible vortexable bands with non-uniform Berry curvatures and FCI ground states. Our finding thus poses an interesting open question as to whether the YCl electride can support FCI states. On the other hand, since the Berry curvature provides a lower bound for the trace of Fubini-Study quantum metric tensors~\cite{Roy2014PRB}, the gate-tunable high Chern number and giant Berry curvatures of order $10^4 {\AA}^2$ near $\pm K$ indicate prominent quantum-metric effects, which bears profound implications for electron correlation physics~\cite{Kitamura2024PRL, Kudo2025arXiv, Peotta2015, Chen2024PRL} worthy of future investigations.\\

\emph{Acknowledgement.} --- The authors thank Adrian H. C. Po, Xi Dai, Tingxin Li, Hongchao Xie, Akito Daido, Jin-Xin Hu, K. T. Law, and Marcel Franz for illuminating discussions. We acknowledge the support of NSFC-Young Scientists Fund (No.12504194, No. 12304093, No. 12447158), Guangdong Provincial Quantum Science Strategic Initiative (Grant No. GDZX2501004), Guangdong Provincial Talents Program 2025D03J0006, Guangzhou Basic and Applied Basic Research Scheme (No. 2024A04J4509), and Start-up Fund of HKUST(GZ) through Grant No. G0101000127 and No. G0101000263. 

\bibliographystyle{apsrev4-2}
\bibliography{main.bib}

\clearpage

\onecolumngrid
\setlength{\textheight}{24cm}
\setlength{\textwidth}{17cm}
\setlength{\parskip}{0pt}

\begin{center}


{\large \textbf{Supplemental Material: YCl Electride as a Multi-Orbital Correlated Topological Dice Lattice System}}\\[1em]

Jianqi Zhong,
Songyuan Geng,
Teng-Fei Ying, 
Haoxiang Li,
Benjamin T. Zhou$^*$\\[0.5em]

{\it
Thrust of Advanced Materials \& Quantum Science and Technology Center,
The Hong Kong University of Science and Technology (Guangzhou), 1 Duxue Rd., Nansha, Guangzhou, China
}\\[0.5em]

\end{center}

\vspace{1em}

\setcounter{section}{0}
\renewcommand{\thesection}{S\arabic{section}}

\setcounter{figure}{0}
\renewcommand{\thefigure}{S\arabic{figure}}

\setcounter{table}{0}
\renewcommand{\thetable}{S\arabic{table}}

\setcounter{equation}{0}
\renewcommand{\theequation}{S\arabic{equation}}

\setlength{\parindent}{2em}
\large

\section{I. DFT calculation for monolayer and multilayer YCl}
First-principles calculations are carried out within the framework of Density Functional Theory (DFT) using the Vienna \textit{Ab initio} Simulation Package (VASP)~\cite{Furthmuller1996,Kresse1996CMS} and the projector augmented-wave (PAW) method~\cite{Vanderbilt1990,KresseHafner1994,Andersen1975}. The exchange-correlation potential is described by the local density approximation (LDA) with the Ceperley-Alder form as parametrized by Perdew and Zunger (CAPZ)~\cite{Ceperley1980,Perdew1981}. A kinetic energy cutoff of 600 eV is applied to the plane-wave basis, which guarantees excellent convergence of results for both total energies and atomic forces. For Brillouin zone sampling, we employ Monkhorst-Pack meshes, typically $12\times12\times1$ for multilayer systems. All structural optimizations, including different magnetic configurations, are conducted until the Hellmann–Feynman forces on atoms are smaller than 0.002 eV/Å and the total energy converges within negligible error of $10^{-6}$ eV. 

\begin{figure}[h]
	\centering
	\includegraphics[width=1\textwidth]{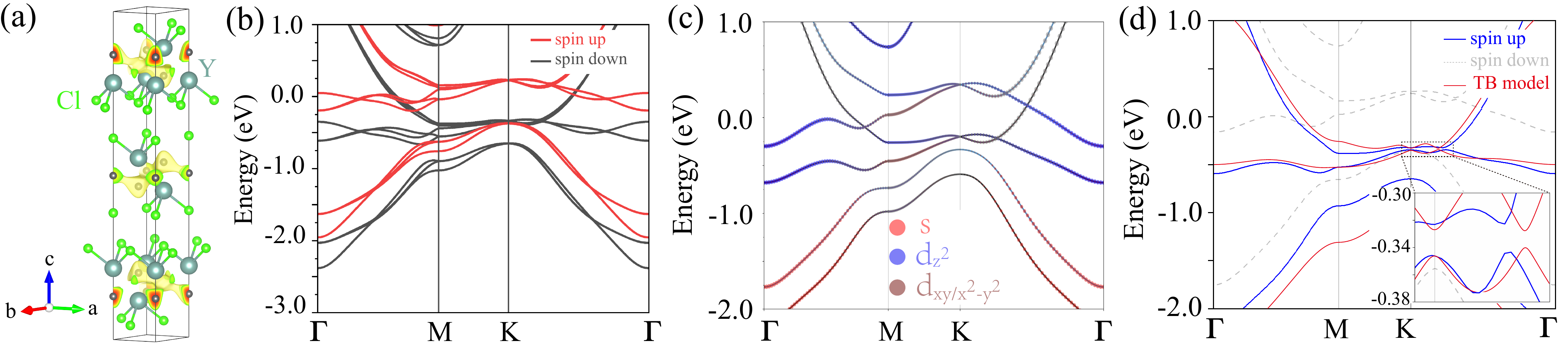}
	\caption{ (a) The distribution of electron localization function (ELF) in the bulk crystal structure. Black dots represent the interstitial effective electron lattice sites (Wannier orbital centers) displaced from the ionic lattice framework, where the electron density is highly concentrated. (b) The ferromagnetic multilayer band structure without SOC. (c) The ferromagnetic monolayer band structure without SOC. (d) The comparisons of DFT calculation and tight-binding model with SOC for monolayer.
	}  
	\label{fig:S1}
\end{figure}

In accordance with the recent ARPES results and DFT calculations in Ref.~\cite{Geng2026NC}, the formation of electride is identified through a detailed analysis of charge localization. We employ an extended Bader charge analysis in combination with the electron localization function (ELF), illustrated in Fig.~\ref{fig:S1} (a). The badelf algorithm is used to trace interstitial electrons and identify electride-like regions, providing a robust way to map excess electron density not bound to atomic sites. 

To investigate the two-dimensional nature of layered YCl, calculations are performed for both multilayer and monolayer structures. A vacuum spacing of 20 {\AA} is introduced along the out-of-plane $c$-axis to eliminate spurious interactions between periodic images. The electronic structure of the monolayer is found to be in good agreement with that of the bulk, as shown by the multilayer versus monolayer YCl band structures shown in Fig.~\ref{fig:S1} (b) and (c). 

Spin-orbit coupling (SOC) is explicitly included in selected calculations to capture relativistic interactions, which are relevant for the fine features of the $d$-bands discussed in the main text. SOC is incorporated using the second-variational method as implemented in VASP, after initial non-SOC relaxations were fully converged.

Last but not least, the edge spectral function presented in Fig.2 of the main text to confirm the bulk-edge correspondence is obtained by the WannierTools~\cite{Wu2018CPC}.

\section{II. The $k \cdot p$ model around $\pm K$ points}

As we mentioned in the main text, there also exists other band crossings along the $K' - \Gamma-K$ line (showing Fig.2a-c in the main text). As the two opposite $K$ valleys are related by inversion $\mathcal{P}$ and the three inequivalent band crossing points at $\bm{k}_{\rm N, 1}, \bm{k}_{\rm N, 2}, \bm{k}_{\rm N, 3}$ around each $K$ are related by $\mathcal{C}_{3z}$, without loss of generality we consider $\bm{k}_{\rm N, 1}$ along the $\Gamma-K$ line. The subgroup for a generic point along the $\Gamma-K$ line contains (i) the (spinless) two-fold rotation along the $x$-axis $\mathcal{C}_{2x} \equiv \mathcal{M}_x \mathcal{P}$, and (ii) the space-time inversion $\mathcal{PT}$. The band crossings occur along the $\mathcal{C}_{2x}$-invariant $\Gamma-K$ line between branches of states with opposite $\mathcal{C}_{2x}$-parities. In fact, near the node at $\bm{k}_{\rm N, 1}$ one can always construct a pseudospin basis $\{\ket{\bm{k}_{N}, +}, \ket{\bm{k}_{N}, -} \}$ which satisfies: $\mathcal{C}_{2x}\ket{\bm{k}_{N}, \pm} = \pm \ket{\bm{k}_{N}, \pm}$ and $\mathcal{PT}\ket{\bm{k}_{N}, \pm} = \pm \ket{\bm{k}_{N}, \pm}$. Under the pseudospin basis, the low-energy Hamiltonian near the node $\bm{k}_{\rm N}$ must satisfy the constraints $\tilde{\tau_z} H_{\rm N}(p_x, p_y) \tilde{\tau_z} = H_{\rm N}(p_x, -p_y)$ and $\tilde{\tau_z} H^{\ast}_{\rm N}(\bm{p}) \tilde{\tau_z} = H_{\rm N}(\bm{p})$, which leads to a massless Dirac fermion:
\begin{equation}\label{eq:masslessDiracN}
H_{N}(\bm{p}) = \tilde{v}_x p_x \tilde{\tau}_z + \tilde{v}_y p_y \tilde{\tau}_y,    
\end{equation}
where $\bm{p} = (p_x, p_y)$ is the momentum measured from $\bm{k}_{\rm N}$, and $\tilde{\tau}_{\alpha=x,y,z}$ are the Pauli matrices for pseudospins. 

\large
\section{III. The seven-band tight-binding model} \label{Section II}
In this section, we construct the seven-band tight-binding model for monolayer YCl under symmetry constraints. The monolayer YCl belongs to the space group $P\bar{3}m1$ (No.~164), corresponding to the $D_{3d}$ point group. For simplicity we consider a single-spin sector as the energy bands of monolayer YCl are spin-polarized. In the non-relativistic limit, there is an extra spinless time-reversal symmetry $\mathcal{T}$, which amounts to complex conjugation, and the full symmetry group of the system is thus given by $D_{3d}\otimes\mathcal{T}$. 

Based on the orbital projection from DFT calculations, the dominant orbitals for the dice bands near the Fermi energy consist of $d_{z^2}$, $d_{xy}$, $d_{x^2-y^2}$ and 5$s$ from the Y atoms. The Bloch basis is constructed from Wannier orbitals $\phi_{l,m}(\mathbf{r}-\mathbf{R}_l)$ on three sublattices $l=t,s,b$. The monolayer YCl has a three-layer lattice architecture as shown in Fig.1 of the main text, with the top ($t$) and bottom ($b$) layers hosting the $d_{z^2}$ and $d_{x^2-y^2}\pm i d_{xy}$ orbitals, while the central sublattice ($s$) accommodates the $s$ orbital. The magnetic quantum numbers are given by $m=0$ for $s$ and $d_{z^2}$, and $m=\pm 2$ for $d_{x^2-y^2}\pm i d_{xy}$. We arrange the Bloch basis of the seven orbitals in the following order: ${\ket{\bm{k},0,t}, \ket{\bm{k},+2,t}, \ket{\bm{k},-2,t}, \ket{\bm{k},0,s}, \ket{\bm{k},0,b}, \ket{\bm{k},+2,b}, \ket{\bm{k},-2,b}}$, and in this basis the single-particle Hamiltonian can be written in blocks:
	\begin{equation}
		H_{0}(\bm{k})=\left[\begin{array}{cccccc}
			h_t(\bm{k}) & h_{ts}(\bm{k}) & h_{tb}(\bm{k})  \\
			h_{ts}^{\dagger}(\bm{k}) & h_s(\bm{k}) &  h_{sb}(\bm{k})  \\
			h_{tb}^{\dagger}(\bm{k})  & h_{sb}^{\dagger}(\bm{k})  & h_b(\bm{k}) \\
		\end{array}\right],
        \label{EqS1}
	\end{equation}
where $h_t(\bm{k})$, $h_b(\bm{k})$ are 3$\times$3 matrices describing the intra-layer terms among the three different $d-$ orbitals in the top and bottom layers. $h_s(\bm{k})$ is a 1$\times$1 matrix describing the excess $s$-electron in the central layer, and $h_{ts}(\bm{k})$ and $h_{sb}(\bm{k})$ are 3$\times$1 and 1$\times$3 matrices respectively for the inter-layer $s$-$d$ hybridization terms, $h_{tb}(\bm{k})$ is a 3$\times$3 matrix describing the inter-layer terms among the three different $d$- orbitals in-between the top and bottom layers. 

The form of the seven-band Hamiltonian is further constrained by the $D_{3d}\otimes\mathcal{T}$ symmetries. For any symmetry operation $\hat{g}\in D_{3d}$, the general symmetry constraint is formulated as:
	\begin{equation}
		U(\hat{g})\, H_{0}(\bm{k})\, U^\dagger(\hat{g}) 
		= H_{0}(\hat{g}\bm{k}) ,
		\label{EqS2}
	\end{equation}
while the spinless time-reversal symmetry $\mathcal{T}$ dictates:
	\begin{equation}
		U_{\mathcal{T}}\, H_{0}(\bm{k})\, U_{\mathcal{T}}^{-1} 
		= H_{0}^*(-\bm{k}) ,
		\label{EqS3}
	\end{equation}
Since Eq.~\ref{EqS2} holds for all elements $\hat{g}\in D_{3d}$ once it is satisfied for all the generators, it suffices to consider three distinct generators: (i) three-fold rotation $C_{3z}$, (ii) in-plane mirror $M_x$, and (iii) spatial inversion $P$. The matrix representations of the three generators in the seven-orbital basis are given by:
	\begin{align}
		U(C_{3z}) &=
		\begin{pmatrix}
			U_d(C_{3z}) & 0_{3\times 1} & 0_{3\times 3} \\
			0_{1\times 3} & 1 & 0_{1\times 3} \\
			0_{3\times 3} & 0_{3\times 1} & U_d(C_{3z})
		\end{pmatrix}, \nonumber \\[6pt]
		U(M_x) &=
		\begin{pmatrix}
			U_d(M_x) & 0_{3\times 1} & 0_{3\times 3} \\
			0_{1\times 3} & 1 & 0_{1\times 3} \\
			0_{3\times 3} & 0_{3\times 1} & U_d(M_x)
		\end{pmatrix}, \nonumber \\[6pt]
		U(P) &=
		\begin{pmatrix}
			0_{3\times 3} & 0_{3\times 1} & I_{3\times 3} \\
			0_{1\times 3} & 1 & 0_{1\times 3} \\
			I_{3\times 3} & 0_{3\times 1} & 0_{3\times 3}
		\end{pmatrix},
		\label{EqS4}
	\end{align}
where the $3\times 3$ blocks acting on the $d$ orbitals are:
	\begin{equation}
		U_d(C_{3z})=
		\begin{pmatrix}
			1 & 0 & 0 \\
			0 & \omega_+ & 0 \\
			0 & 0 & \omega_-
		\end{pmatrix},\qquad
		U_d(M_x)=
		\begin{pmatrix}
			1 & 0 & 0 \\
			0 & 0 & 1 \\
			0 & 1 & 0
		\end{pmatrix},
		\label{EqS5}
	\end{equation}
with $\omega_\pm \equiv e^{\pm i 2\pi/3}$ the nontrivial eigenvalues of $C_{3z}$. We note that $|d_{x^2-y^2}\rangle \pm i |d_{xy}\rangle$ form time-reversal partners, thus the representation of $\mathcal{T}$ is identical to that of $M_x$, i.e., $U_{\mathcal{T}} = U(M_x)$.

Imposing Eqs.~\ref{EqS3},~\ref{EqS4} on $H_{TB}(\bm{k})$ in Eq.~\ref{EqS2} and taking into account terms up to next-nearest-neighbor hopping, we 
work out the final form of $H_{TB}(\bm{k})$ as:
\begin{equation}
\begin{aligned}
& h_t(\boldsymbol{k})= \\
& \left(\begin{array}{ccc}
E_{d 0}+2 t_{d 0} C(\boldsymbol{k}) & 2 t_{d 02} C_{-}(\boldsymbol{k}) & 2 t_{d 02} C_{+}(\boldsymbol{k}) \\
2 t_{d 02} C_{+}(\boldsymbol{k}) & E_{d 2}+2 t_{d 2} C(\boldsymbol{k})+2 u_{d 2} S(\boldsymbol{k}) & 2 t_{d 22} C_{-}(\boldsymbol{k}) \\
2 t_{d 02} C_{-}(\boldsymbol{k}) & 2 t_{d 22} C_{+}(\boldsymbol{k}) & E_{d 2}+2 t_{d 2} C(\boldsymbol{k})-2 u_{d 2} S(\boldsymbol{k})
\end{array}\right), \\
& h_b(\boldsymbol{k})= \\
& \left(\begin{array}{ccc}
E_{d 0}+2 t_{d 0} C(\boldsymbol{k}) & 2 t_{d 02} C_{-}(\boldsymbol{k}) & 2 t_{d 02} C_{+}(\boldsymbol{k}) \\
2 t_{d 02} C_{+}(\boldsymbol{k}) & E_{d 2}+2 t_{d 2} C(\boldsymbol{k})-2 u_{d 2} S(\boldsymbol{k}) & 2 t_{d 22} C_{-}(\boldsymbol{k}) \\
2 t_{d 02} C_{-}(\boldsymbol{k}) & 2 t_{d 22} C_{+}(\boldsymbol{k}) & E_{d 2}+2 t_{d 2} C(\boldsymbol{k})+2 u_{d 2} S(\boldsymbol{k})
\end{array}\right), \\
& h_s(\boldsymbol{k})=E_s+2 t_s C(\boldsymbol{k}), \\
& h_{t s}(\boldsymbol{k})=\left(\begin{array}{l}
\gamma_0 f_0(\boldsymbol{k}) \\
\gamma_2 f_{+}(\boldsymbol{k}) \\
\gamma_2 f_{-}(\boldsymbol{k})
\end{array}\right), h_{s b}(\boldsymbol{k}) =\left(\begin{array}{lll}
\gamma_0 f_0(\boldsymbol{k}) & \gamma_2 f_{-}(\boldsymbol{k}) & \left.\gamma_2 f_{+}(\boldsymbol{k})\right),
\end{array}\right. \\
& h_{t b}(\boldsymbol{k})  =\left(\begin{array}{lll}
g_0 f_0(-\boldsymbol{k}) & \left.g_1 f_{-}-\boldsymbol{k}\right) & \left.g_1 f_{+}-\boldsymbol{k}\right) \\
g_1 f_{+}(-\boldsymbol{k}) & g_2 f_0(-\boldsymbol{k}) & g_3 f_{-}(-\boldsymbol{k}) \\
g_1 f_{-}(-\boldsymbol{k}) & g_3 f_{+}(-\boldsymbol{k}) & g_2 f_0(-\boldsymbol{k})
\end{array}\right).
\end{aligned}
\end{equation}
where $C(\bm{k}) = \sum_{j=1}^{3} \cos(\bm{k}\cdot \bm{R}_j), 
C_{\pm}(\bm{k}) = \sum_{j=1}^{3} \omega_{\pm}^{\,j-1} \cos(\bm{k}\cdot \bm{R}_j), 
S(\bm{k}) = \sum_{j=1}^{3} \sin(\bm{k}\cdot \bm{R}_j), 
f_0(\bm{k}) = \sum_{j=1}^{3} e^{i\bm{k}\cdot \bm{\delta}_j}, $ and $f_{\pm}(\bm{k}) = \sum_{j=1}^{3} \omega_{\pm}^{\,j-1} e^{i\bm{k}\cdot \bm{\delta}_j}$ with the unit lattice vectors
$\bm{R}_1=(a,0)$, $\bm{R}_2=(-a/2,\sqrt{3}a/2)$, $\bm{R}_3=(-a/2,-\sqrt{3}a/2)$,  
and the nearest-neighbor bonding vectors $\bm{\delta}_1=\tfrac{a}{\sqrt{3}}(0,1)$,
$\bm{\delta}_2=\tfrac{a}{\sqrt{3}}(-\sqrt{3}/2,-1/2)$,
$\bm{\delta}_3=\tfrac{a}{\sqrt{3}}(\sqrt{3}/2,-1/2)$.  
The TB parameters are fitted from the first-principle band structures obtained by DFT+U calculation and summarized in Table ~\ref{table1} below. The band structures of the TB model with and without atomic SOC are shown in Fig.2 of the main text.

Next, we present details of the Wilson-loop method used for calculating the Chern number. Given the eigenvalue equation:
\begin{equation}
    H_{0}(k_x,k_y) \ket{u_n(k_x, k_y)}= E_n(k_x,k_y) \ket{u_n(k_x, k_y)},
\end{equation}
we define the following overlap matrix along $y$-direction occupied bands:
\begin{equation}
    M_y^{mn}(k_x;k_y^j, k_y^{j+1})= \braket{u_m(k_x, k_y^j)|u_n(k_x, k_y^{j+1})},
\end{equation}
where $k_y^j=2\pi j/N, m, n =1, 2,.., n_{occ}$. The Wilson loop is then written as:
\begin{equation}
   \omega_y (k_x)=\prod \limits_{j=0}^{N-1} M_y^{mn}(k_x;k_y^j,  k_y^{j+1}),
\end{equation}
Straightforward algebraic operations lead to $\omega_y (k_x)=e^{-i\oint_{-\pi }^{\pi }A_{y}^{mn}(k_x,k_y)dk_y }$, where $A_{y}^{mn}(k_x,k_y)=-i \bra{u_m^k} \partial_{k_y} \ket{u_n^k}$. This allows us to obtain the polarization along $y$-direction: $p_y(k_x)=-\frac{i}{2\pi }\ln \textrm{det}[\omega_y(k_x)]
=-\frac{1}{2 \pi} \int_0^{2 \pi}$Tr$[A_y(\bm k)] dk_y$, and the Chern number $\mathcal{C}$ is given by:
\begin{equation}
\mathcal{C}=\int_0^{2 \pi} dk_x \partial_{k_x} p_y(k_x).    
\end{equation}

\begin{table}[h]
    \centering
    \caption{Seven-band tight-binding parameters (eV per unit) obtained by fitting DFT+U results.}
    \label{tab3}
    \renewcommand\arraystretch{1.4}
    \large
    \begin{tabular}{lcccccccccccccccc}
        \hline\hline
        \rule{0pt}{13pt}
        & $E_s$  & $E_{d0}$ & $E_{d2}$ & $t_s$ & $t_{d0}$ & $t_{d2}$ & $u_{d2}$ & $t_{d02}$ & $t_{d22}$ & $\gamma_0$ & $\gamma_2$ & $g_0$ & $g_1$ & $g_2$ & $g_3$ \\
        \hline
        & -0.5 & 0.25 & 1.0 & -0.01 & -0.02 & 0.1 & -0.1 & 0.08 & 0.12 & -0.4 & -0.01 & 0.06 & -0.04 & -0.01 & -0.14 \\
        \hline\hline
    \end{tabular}
    \label{table1}
\end{table}

\section{IV. The self-consistent Hartree-Fock calculations}

In this section, we study interaction-driven magnetism in a seven-band tight-binding model $H_0$ (see the main text). For compactness, we denote a composite index $\alpha=(m,\sigma)$ where $m$ label orbitals within one sublattice site and $\sigma\in\{\uparrow,\downarrow\}$ denotes spin. The interacting Hamiltonian reads
\begin{equation}
\hat H_{\mathrm{int}} =\sum_{i,m} U_m \,\hat n_{i m \uparrow} \hat n_{i m \downarrow} +
\sum_{i,m\neq m'} \sum_{\sigma,\sigma'} U'_{mm'}(\sigma,\sigma') \, \hat n_{i m \sigma} \hat n_{i m' \sigma'},
\label{S11}
\end{equation} 
where we assume that 
\begin{equation}
	U'_{mm'}(\sigma,\sigma')=
	\begin{cases}
		U' - J, & \sigma = \sigma', \\
		U',     & \sigma \ne \sigma'.
	\end{cases}
	\label{S12}
\end{equation}
here, $U_m$ for the first term of Eq.\ref{S11} and $U^\prime $ are the intra- and inter-orbital interacting, respectively. After performing the Fourier transformation $c_{i,m\sigma}=N^{1/2}\sum_k e^{i \boldsymbol{k} \cdot \boldsymbol{R_i}} c_{i,\boldsymbol{k},m\sigma}$, where we also label the site index $i \in $ (A or B) to constrain the site-dependent inter-orbital interactions for the second term of Eq.\ref{S11}. Thus, Eq.\ref{S11} in momentum space can be simplified as:
\begin{equation}
\begin{aligned}
\hat H_{\mathrm{int}}= &\;\frac{1}{N}\sum_{m}\sum_{\mathbf k,\mathbf k',\mathbf q}U_m \,
c^{\dagger}_{\mathbf k+\mathbf q,m\uparrow}
c_{\mathbf k,m\uparrow}
c^{\dagger}_{\mathbf k'-\mathbf q,m\downarrow}
c_{\mathbf k',m\downarrow} \\
&+\frac{1}{N}\sum_{i,m\neq m'}\sum_{\sigma,\sigma'}\sum_{\mathbf k,\mathbf k',\mathbf q}U'_{i,mm'}(\sigma,\sigma')\,
c^{\dagger}_{\mathbf k+\mathbf q,i,m\sigma}
c_{\mathbf k,i,m\sigma}
c^{\dagger}_{\mathbf k'-\mathbf q,i,m'\sigma'}
c_{\mathbf k',i,m'\sigma'} .
\end{aligned}
\label{S13}
\end{equation}

Within the mean-field method, by defining the density matrix $P^i_{\alpha, \beta}(\mathbf{k})=\left\langle c_{i,\mathbf{k}, \alpha}^{\dagger} c_{i,\mathbf{k}, \beta}\right\rangle$, the Hartree term is diagonal in orbital and spin indices:
\begin{equation}
	\begin{aligned}
		h_{U, H} & =\frac{U}{N} \sum_{\mathbf{k}, m, \sigma}\left[\sum_{\mathbf{k}^{\prime},\bar{\sigma}} P_{(m, \bar{\sigma})(m, \bar{\sigma})}\left(\mathbf{k}^{\prime}\right)\right] c_{\mathbf{k}, m, \sigma}^{\dagger} c_{\mathbf{k}, m, \sigma} \\
		h_{U^{\prime}, H} & =\frac{U^\prime}{N} \sum_{\mathbf{k}, i,m \neq m^{\prime}, \sigma}\left[ \sum_{\mathbf{k}^{\prime}} P^i_{\left(m^{\prime}, \sigma\right)\left(m^{\prime}, \sigma\right)}\left(\mathbf{k}^{\prime}\right)+ \sum_{\mathbf{k}^{\prime},\bar{\sigma}} P^i_{\left(m^{\prime}, \bar{\sigma}\right)\left(m^{\prime}, \bar{\sigma}\right)}\left(\mathbf{k}^{\prime}\right)\right] c_{\mathbf{k}, i, m, \sigma}^{\dagger} c_{\mathbf{k}, i, m, \sigma}, \\
        h_{J,H} &= -\frac{J}{N} \sum_{\mathbf{k}, i,m \neq m^{\prime}, \sigma} \left[\sum_{\mathbf{k}^{\prime}} P^i_{\left(m^{\prime}, \sigma\right)\left(m^{\prime}, \sigma\right)}\left(\mathbf{k}^{\prime}\right) \right] c_{\mathbf{k},i, m, \sigma}^{\dagger} c_{\mathbf{k},i, m, \sigma},
	\end{aligned}
    \label{S14}
\end{equation}
The Fock contributions arise from exchange interactions and are generally off-diagonal in spin (and potentially orbital) indices. Under the density-density interaction considered in this work, the Fock term reads
\begin{equation} \label{S15}
\begin{aligned}
    h_{U, F}&=-\frac{U}{N} \sum_{\mathbf{k}, m}\left[\sum_{\mathbf{k}^{\prime}} P_{(m, \uparrow)(m, \downarrow)}\left(\mathbf{k}^{\prime}\right)\right] c_{\mathbf{k}, m, \downarrow}^{\dagger} c_{\mathbf{k}, m, \uparrow}, \\
    h_{U^\prime, F}&=-\frac{U^\prime}{N} \sum_{\mathbf{k}, i,m \neq m^{\prime}}\left[\sum_{\mathbf{k}^{\prime}} P^i_{\left(m^{\prime}, \uparrow\right)\left(m, \uparrow\right)}\left(\mathbf{k}^{\prime}\right) c_{\mathbf{k},i, m, \uparrow}^{\dagger} c_{\mathbf{k},i, m^\prime, \uparrow}+\sum_{\mathbf{k}^{\prime}} P^i_{(m^\prime, \uparrow)(m, \downarrow)}\left(\mathbf{k}^{\prime}\right) c_{\mathbf{k}, i,m, \downarrow}^{\dagger} c_{\mathbf{k},i, m^\prime, \uparrow}\right], \\
    h_{J,F} &= \frac{J}{N} \sum_{\mathbf{k},i, m \neq m^{\prime}} \left[\sum_{\mathbf{k}^{\prime}} P^i_{\left(m^{\prime}, \uparrow\right)\left(m, \uparrow\right)}\left(\mathbf{k}^{\prime}\right) \right] c_{\mathbf{k}, i,m, \uparrow}^{\dagger} c_{\mathbf{k},i, m^\prime, \uparrow}.
\end{aligned}
\end{equation}
Starting from an initial density matrix $P_{n=0}$, we construct the corresponding Hartree-Fock (HF) Hamiltonian $h_n^{\mathrm{HF}}$ using Eq.\ref{S14} and \ref{S15}, where $n$ denotes the iteration steps.
To obtain convergent mean-field solutions, we construct the initial density matrix in two steps. 
We first consider the on-site Hubbard model with only intra-orbital interaction and introduce a small spin polarization along the $z$ direction (proportional to $\sigma_z$) as the initial perturbation. 
This procedure yields a self-consistent anti-ferromagnetic (AFM) solution for $\nu=4$. The resulting AFM density matrix is then used as the initial guess for the full multi-orbital Hubbard calculation, which can help us get the lowest ground state. To ensure numerical stability of the self-consistent iteration, we also introduce a damping factor $\alpha$ and employ a linear mixing scheme between successive iterations,
say $P_{n+1}^{\mathrm{in}} = (1-\alpha) P_n^{\mathrm{in}} + \alpha P_{n+1}^{\mathrm{out}}$, and set $\alpha=0.5$. The maximum number of iterations is set by $n_{\rm max}=2\times$10$^4$. The calculations are performed at finite temperature of $T\approx 10$K with $\beta = 1/(k_B T) = 10^3$, and a convergence criterion of $\delta P=|P_{n+1}^{\mathrm{in}}-P_{n}^{\mathrm{in}} |<10^{-8}$ is imposed.

\begin{figure}[h]
	\centering
	\includegraphics[width=1.0\textwidth]{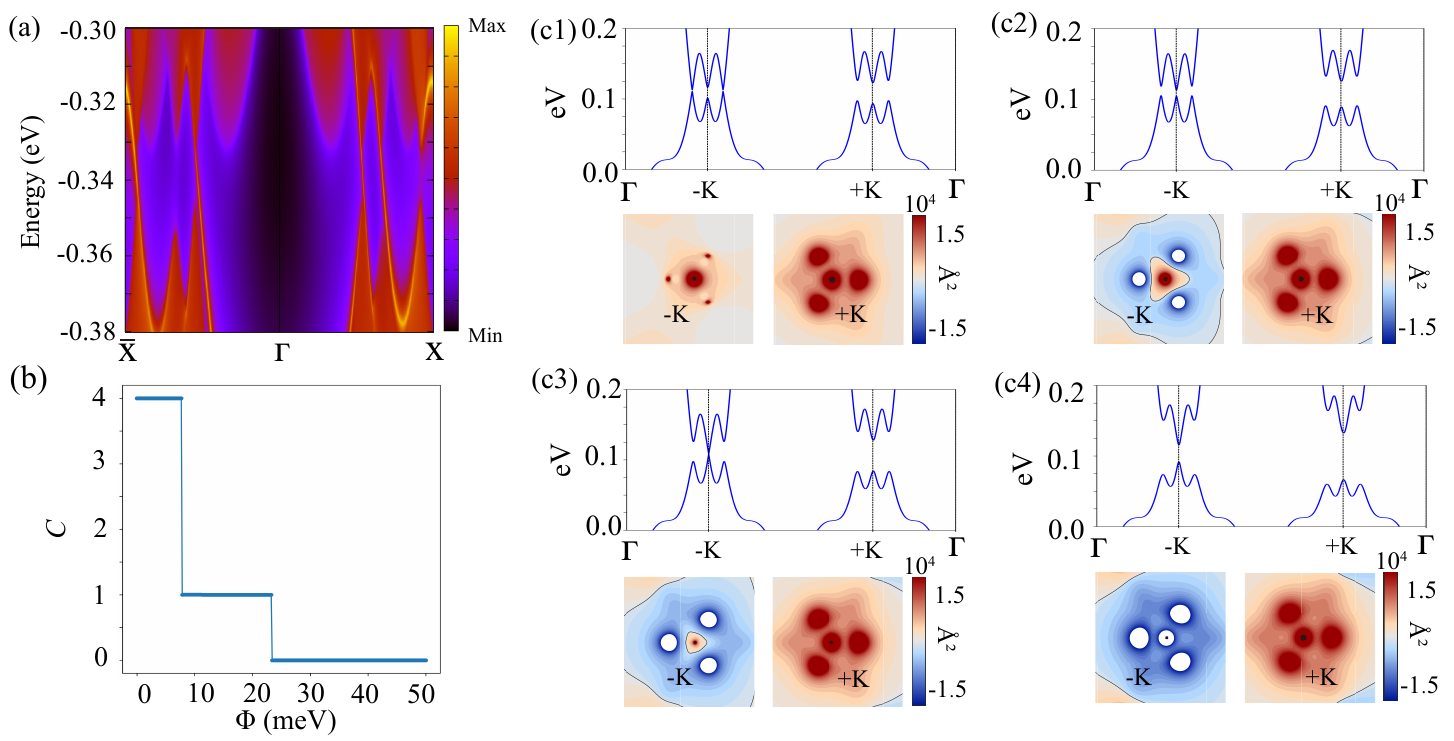}
	\caption{(a) The edge spectral function plot in logarithmic scale obtained for an infinite strip geometry, signifying 4 branches of chiral edge states (bright yellow curves). (b) Chern number as a function of $\Phi$ for $\lambda=15$meV. 
    (c1)-(c4) Evolution of energy bands and Berry curvatures under increasing inter-layer potential difference $\Phi$, where  $\Phi_{\rm c1}=7.7$meV, $\Phi_1=15$meV, $\Phi_{\rm c2}=23.2$meV, $\Phi_2=50$meV, respectively. The bottom panel shows the corresponding Berry curvature distribution near $\pm K$ in each case. 
	}  
	\label{Sfig3}
\end{figure}

\section{V. Evolution of topological transitions without/with interactions}

\begin{figure}[h]
	\centering
	\includegraphics[width=1.0\textwidth]{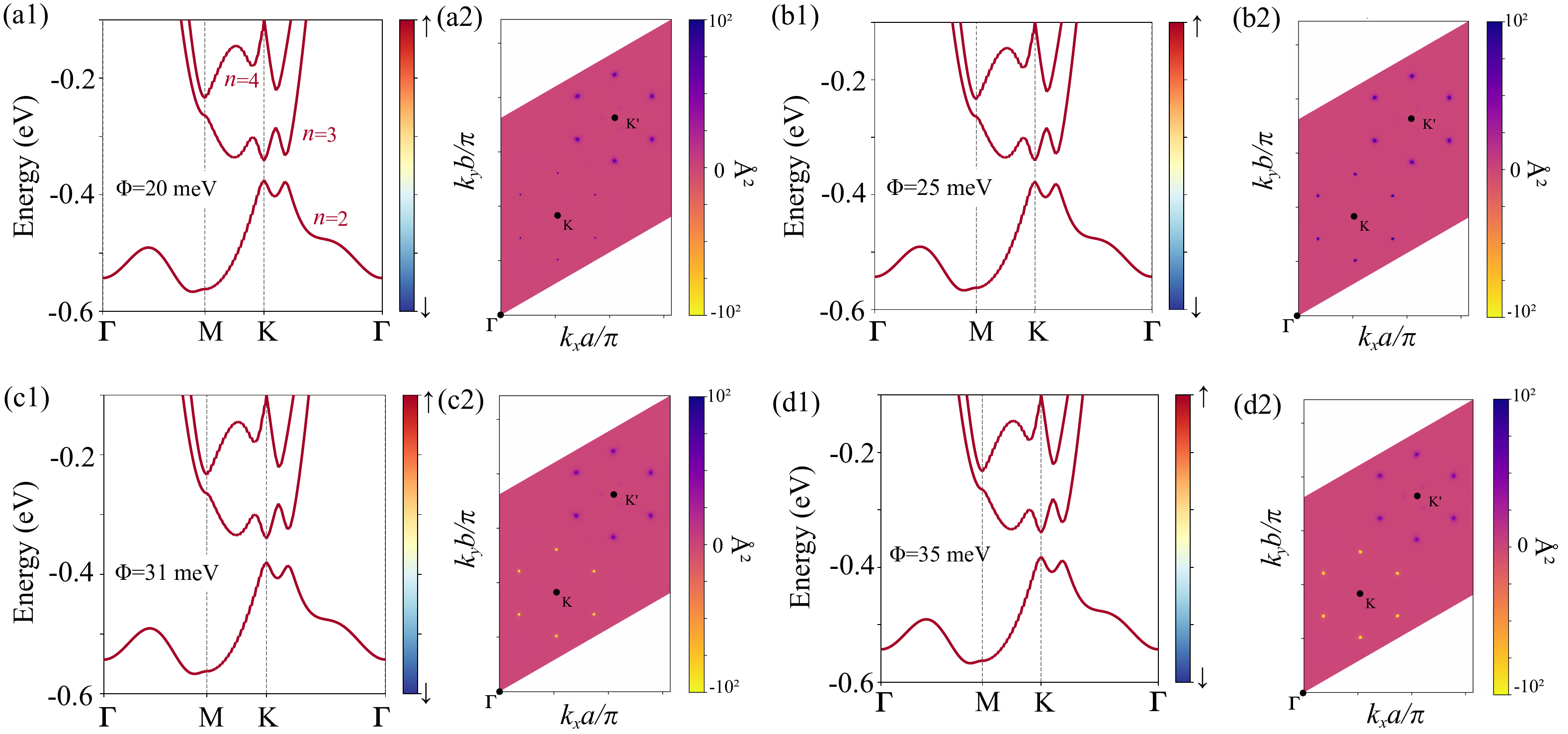}
	\caption{(a) - (d) Evolution of HF band structures and Berry curvatures under increasing displacement field $\Phi$ for $\nu=4$.
	}  
	\label{Sfig3}
\end{figure}

To study the $D_z$-driven topological transition, we introduce $\Phi$ in our realistic tight-binding model with fixed atomic SOC and investigate the band evolution as a function of $\Phi$ (Fig.~\ref{Sfig3}b) including. 

In the non-interacting case, the system remains in the $\mathcal{C} = 4$ phase at a small but finite $\Phi< 10$ meV. Topological gaps at $\bm{k}_{\rm N}$ close and reopen at the first critical $\Phi_{\rm c1} \approx 10$ meV (Fig.~\ref{Sfig3}c1), where the system enters a new topological phase with $\mathcal{C} = 1$ given by the total 2$\pi$ Berry phase from $\pm K$ points. This new $\mathcal{C} = 1$ phase under finite $\Phi$ is closely analogous to the well-celebrated Haldane phase, where $\pi$ Berry phases of the same sign emerge from two different $K$ valleys~\cite{Haldane1988}. As $\Phi$ increases further, gaps at $\pm K$ close a second time at $\Phi_{\rm c2}= 23.2 $ meV (Fig.~\ref{Sfig3}c3,c4), where the system becomes trivial ($\mathcal{C} = 0$).

In the interacting case, some additional topological transitions occur due to band inversions involving the flat band ($n=2$), the upper band ($n=3$), and the higher band ($n=4$) in the spin-up channel (see Fig.~\ref{Sfig3}a). In particular, near the second drop in $\sigma_{xy}$ (as shown in the yellow curve of Fig.3d in the main text), the $n=4$ band mixes with the $n=3$ band and dramatically changes the Berry-curvature distribution as $\Phi$ increases for $\nu=4$. As a result, the net Berry curvature changes sign in the range $\Phi \approx 25$–$31$ meV as shown in Fig~\ref{Sfig3}b1-c2, leading to a strong suppression of $\sigma_{xy}$ toward zero.

\end{document}